# Investigation of the Differential Rotation by Hα Filaments and Long-Lived Magnetic Features for Solar Activity Cycles 20 and 21

**Gigolashvili M. Sh., Japaridze D. R., Kukhianidze V. J.**

E.K. Kharadze Abastumani Astrophysical Observatory at Ilia State University, Tbilisi, Georgia

**Abstract.** For solar activity cycles 20 and 21 (1966-1985) the solar differential rotation has been investigated using Hα filaments and relatively small-scale long-lived magnetic features with negative and positive polarities. We used annual averaged angular velocities of quiescent Hα filaments from Hα photoheliograms of the Abastumani Astrophysical Observatory film collection and selected long-lived magnetic features from the McIntosh atlas (McIntosh, Willock, and Thompson, Atlas of Stackplots, NGDC, 1991). We have determined coefficients of Faye's formulas for Hα filaments as well as for long-lived magnetic features and have found that for solar cycles 20 and 21 the Hα filaments have lower rotation rates and rotated more differentially than the long-lived magnetic features.

## 1. Introduction

Despite of a large amount of various observational data and theoretical investigations the solar differential rotation is not still fully understood. The investigations of the solar differential rotation are presented in reviews (Schröter, 1985; Beck, 2000; Stix, 2002; Rozelot and Neiner, 2009), as well as in several research papers (e.g., Howard, 1976; Gilman and DeLuca, 1989; Howard, Gupta, and Sivaraman, 1999).

The phenomenon of the solar differential rotation has been known for centuries but it is still not properly understood. The measured rotation velocities differ from each other not only for different solar tracer' location in different layers of solar atmosphere, but even for the same type of features.

Since sunspots were recognized as solar features they have been used as tracers for solar rotation (Scheiner, 1630). Besides sunspots, other features of the solar atmosphere used as tracers of solar rotation are e.g., faculae (Newbegin, and Newton, 1931) and Hα filaments (D'Azambuja and D'Azambuja, 1948; Japaridze and Gigolashvili, 1992; Gigolashvili et al., 1995; Gigolashvili et al., 2003, 2005, 2007; Gigolashvili, Japaridze, and Kukhianidze, 2005).

D'Azambuja and D'Azambuja (1948) and Bruzek (1961) were the first to measure rotation rates near the poles using filaments as tracers. Then Adams and Tang (1977) measured the rotation of short-lived Hα filaments as a function of their latitude. Using filaments as tracers Bruzek (1961), Mouradian et al., (1987) and Brajša et al. (1991) found limited solar areas of rigid rotation so-called "pivot points" around which the filaments rotated during two or more successive solar rotations. The association between radio sources and pivot points was analyzed by Brajsa et al. (2000) who also investigated their influence on the measured rotation velocity. Two large stable solar filaments were used as test tracers to determine the apparent synodic rotation rate as a function of the central meridian distance for several filaments' segments of different heights (Vršnak et al., 1999). In this procedure, a method for the simultaneous determination of the solar synodic rotation velocity and the height of the tracers was applied (Roša et al., 1998). The velocity field was derived for many filaments during a few consecutive days and for different Carrington rotations (Ambrož and Schroll, 2002).

Another class of features is the large-scale and small-scale magnetic patterns used for the study of motion and evolution of solar magnetic fields (McIntosh, 1992; Durrant, Turner, and Wilson, 2002;

Gigolashvili, Japaridze, and Kukhianidze, 2005/2006; Japaridze, Gigolashvili, and Kukhianidze 2006, 2007, 2009; Chu et al., 2010, Benevolenskaya, 2010).

The radio images at 17 GHz (Chandra, Vats, and Iyer, 2009) and X-ray bright points over the solar disk (Kariyappa, 2008; Hara, 2009) are used as tracers to determine the coronal rotation as well.

A large amount of observational data and proper theoretical investigations concerning differential rotation and large-scale meridional circulations exist, but no clear relationship between them has been revealed so far. Existence of a close connection between tracers and magnetic fields is a clear fact but the mechanism of their cyclic behaviors by dynamo models has not been explained until today. It was found that magnetic features (sunspots, plages, faculae, etc.) rotate with a higher rotation rate than the solar plasma (e. g. Zaatri et al., 2009).

The aim of present work is to carry out an investigation of solar differential rotation using Hα filaments (HF) long-lived magnetic features (LLMF) of large-scale magnetic fields for Solar Cycles 20 and 21.

## 2. Observational data

We investigate the properties of solar differential rotation by using annually averaged angular velocities of Hα filaments from Hα photoheliograms of the film collection of Abastumani Astrophysical Observatory for the Solar Cycles 20 and 21 (1966-1985). Relatively stable filaments were selected, which did not significantly change their shapes, allowing identification on consecutive days. Hα filaments with lifetimes less than three days had not been measured because precise measurements of the differential rotation rates need at least three independent measurements of coordinates on subsequent days.

We have investigated also the long-lived magnetic features with negative and positive polarity using synoptic charts of the McIntosh atlas (McIntosh, Willock, and Thompson, 1991).

McIntosh's stackplots for the entire range of data for solar cycles 20-21 (1966-1986) include a series of plots displaying 10°-zones of solar latitude, stepped from 60° N through the solar equator up to 60° S. Five identical plots have been placed side-by-side, each stepped up by one row and displaced to the left. Grids to measure rotation rates of the drift patterns accompany the plots (Figure 1).

Segmentation of the charts into stackplots with narrow latitude zones is a valuable method of isolating the differential rotation of the Sun. This differential rotation causes long-lived features at the same latitude to move relatively to those at the adjacent latitude, resulting in complicated interactions among large-scale patterns (McIntosh, Willock, and Thompson, 1991).

## 3. Method of Data Analysis

The differential rotation rates have been measured for Hα filaments and long-lived magnetic features with negative and positive polarities separately for every 10° zones (from 50° N to 50° S) of both hemispheres (Japaridze and Gigolashvili, 1992; Japaridze, Gigolashvili, and Kukhianidze, 2006; 2009).

For measuring the differential rotation rates of Hα filaments we used the standard grids for determining the heliographic longitude and latitude of the features on the solar surface.

The coordinates of well identified points of each Hα filament (with an interval 10°) have been measured for subsequent days and the rotation velocities of Hα filaments have been computed. For reducing errors in estimation of rotation rates we calculated Hα filaments and various fragments of filaments around the solar central meridian in a range ±60° along both longitudes and latitudes.

The solar differential rotation can be estimated from the velocities on the solar disk using a standard approximation:

$$\omega(\phi) = A + B\sin^2\phi + C\sin^4\phi, \tag{1}$$

where $\omega(\phi)$ is the angular velocity in degrees per day, and $\phi$ is the heliographic latitude. $A, B$, and $C$ are solar rotation parameters, in particular $A$ is the equatorial angular velocity and $B, C$ are the latitude gradient coefficients of the rotation rate, $B$ represents mainly low latitudes and $C$ represents mostly higher latitudes. Function (1) defines the differential rotation profile. We can restrict ourselves to only the first two terms of the expansion (1), and set $C = 0$. The result represents the sidereal rotation velocity in deg day$^{-1}$.

To study the differential rotation of long-lived and relatively compact magnetic features among large-scale stackplots patterns we used the atlas of synoptic maps from McIntosh, Willock, and Thompson (1991, pages 175-188).

For determination of rotation rate of long-lived magnetic features, we have developed the following method: we measured the angle between the symmetry axis of a selected magnetic features and the horizontal line parallel to the horizontal edge (Figure 1). The average slope of long-lasting patterns generally varies in a regular way as a function of latitude. Since the frame of reference is the Carrington system of solar longitudes, a vertical pattern in a stackplot represents a pattern rotating at the Carrington synodic rate of 27.2753 days. Patterns with positive or negative slopes indicate rotation rates that are, respectively, faster or slower than the Carrington rate. Rotation rates faster than the Carrington rate usually occur at latitudes less than 20° (McIntosh, Willock, and Thompson, 1991).

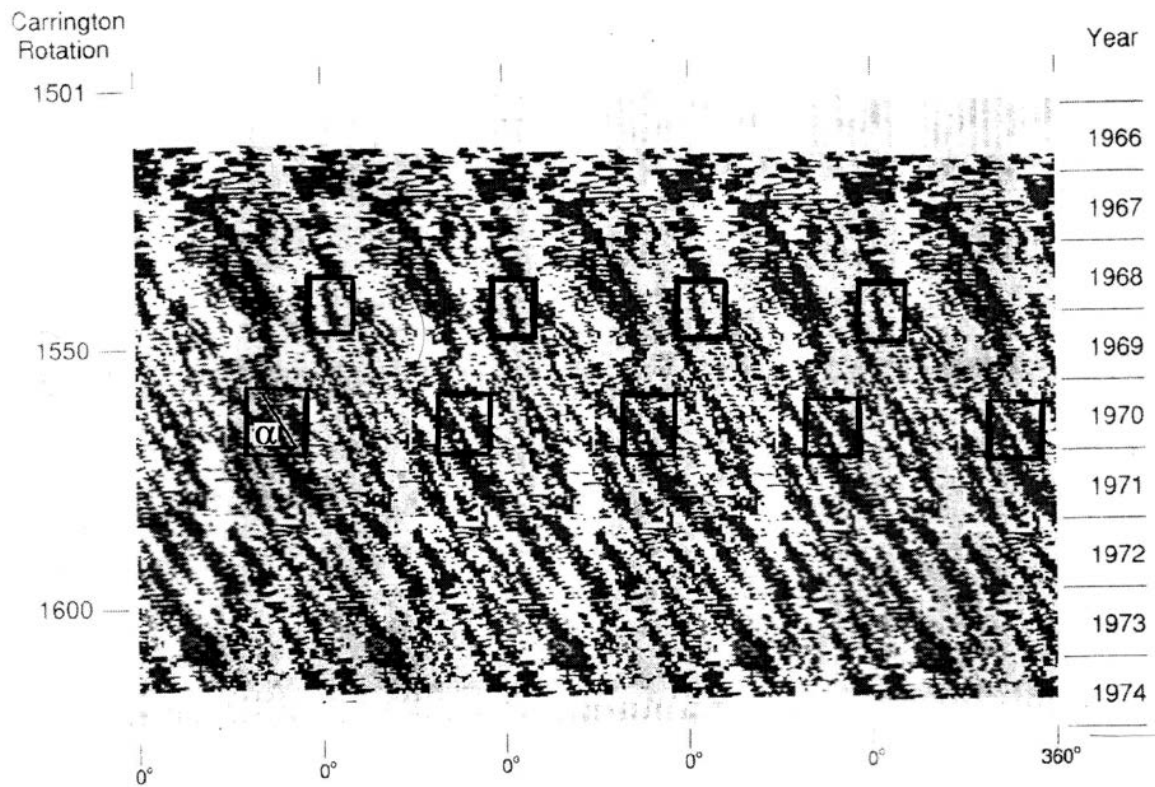

Figure 1. Large-scale stackplot: south 1° to south 10°, for 1966–1974 (page 182). An example of 5 identical area of McIntosh's stackplot is presented. Marked features within the boxes indicate patterns selected by us for measurement: black patterns correspond to negative magnetic features and white patterns to positive magnetic ones. α is the angle between the symmetry axis of a selected magnetic features and the horizontal line parallel to the horizontal edge. The long-lived structural patterns of large-scale magnetic field are presented within the boxes, which are labeled in the five identical plots placed side-by-side.

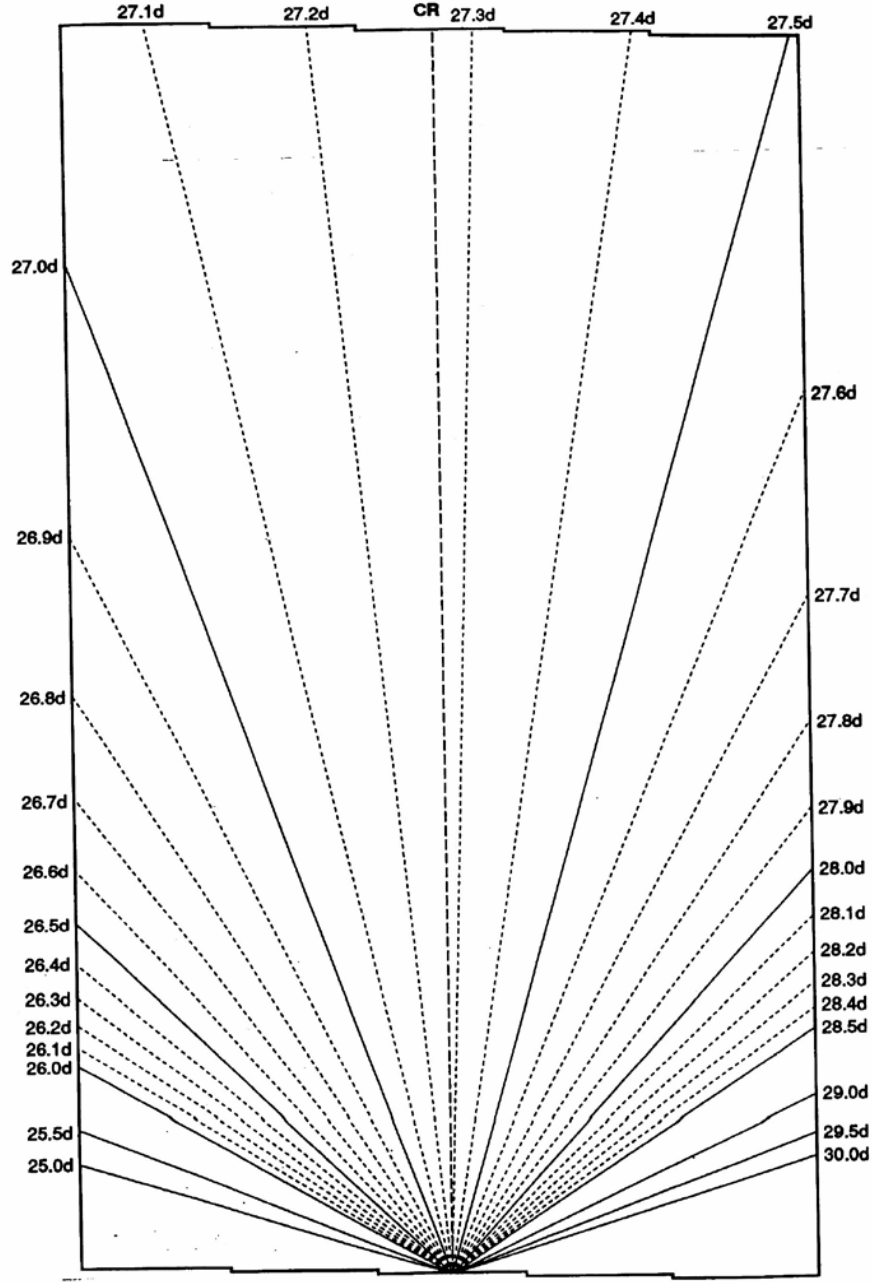

Figure 2. Grids for determination of rotation rate for large-scale stackplots adapted from McIntosh, Willock, and Thompson (1991).

Instead of using the grid (Figure 2), given by McIntosh, Willock, and Thompson (1991), we calculated the rotation rate for given magnetic features by an empirical formula that describes and replaces a grid (Japaridze, Gigolashvili, and Kukhianidze, 2006, 2007):

$$\omega(\varphi) = 1000 / (36.664 - \cot\alpha); \qquad (2)$$

where $\alpha$ represents the angle measured by us for each feature (Figure 1), $\varphi$ is a latitude and $\omega$ is the rotation rate $\omega(\varphi)$, measured in deg day$^{-1}$. This empirical formula has been deduced with means of calculation of values of the rotation rates of some long-lived magnetic features on the different latitudes defined by means of a grid (Figure 2). To select the magnetic features for measuring differential rotation with we have chosen visually symmetric structural formations from a large set of magnetic data (Figure 1).

## 4. Results

We have obtained approximately 11 500 measurements of 517 Hα filaments. Using the McIntosh atlas, we have carried out 1675 measurements for 335 long-lived magnetic features (Table 1).

Table 1. Number of the measurements of all patterns: the various fragments of Hα filaments, long-lived magnetic features with negative and positive polarities, presented separately for Solar northern and southern hemispheres, as well as for full solar disk.

| Cycle Number | Solar Hemisphere | HF | HF Measurements | -LLMF | -LLMF Measurements | +LLMF | +LLMF Measurements |
|---|---|---|---|---|---|---|---|
| 20 | N | 134 | 3000 | 44 | 230 | 46 | 245 |
| 20 | S | 122 | 2500 | 58 | 265 | 49 | 250 |
| 20 | N-S | 256 | 5500 | 102 | 495 | 96 | 495 |
| 21 | N | 130 | 2900 | 38 | 175 | 35 | 169 |
| 21 | S | 131 | 3100 | 35 | 181 | 30 | 160 |
| 21 | N-S | 261 | 6000 | 73 | 356 | 64 | 329 |
| 20-21 | N | 264 | 5900 | 82 | 405 | 81 | 414 |
| 20-21 | S | 253 | 5600 | 93 | 446 | 79 | 410 |
| 20-21 | N-S | 517 | 11500 | 175 | 851 | 160 | 824 |

The diagrams for every 10°-zones were constructed by using average yearly values of the rotation rates for Hα filaments and long-lived magnetic features on the figures 3-4. The rotation rates of Hα filaments and long-lived magnetic features are constructed separately for each year of the Solar Cycles 20 and 21. In these figures (-LLMF) and (+LLMF) are rotation rates of magnetic features with negative and positive polarities, respectively, (HF) is rotation rates of Hα filaments. The error bars of standard deviation are also shown.

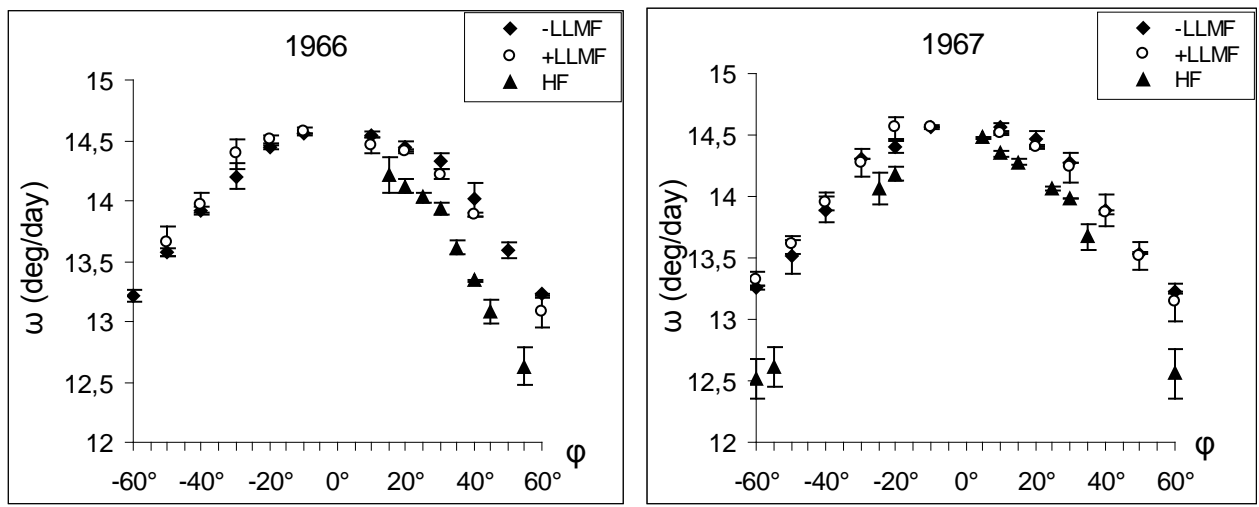

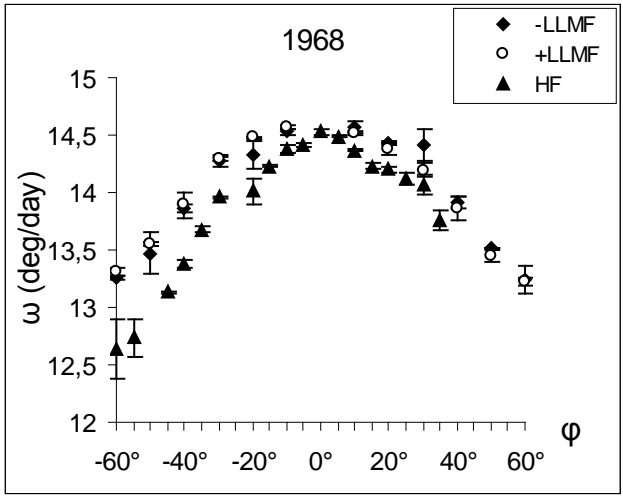
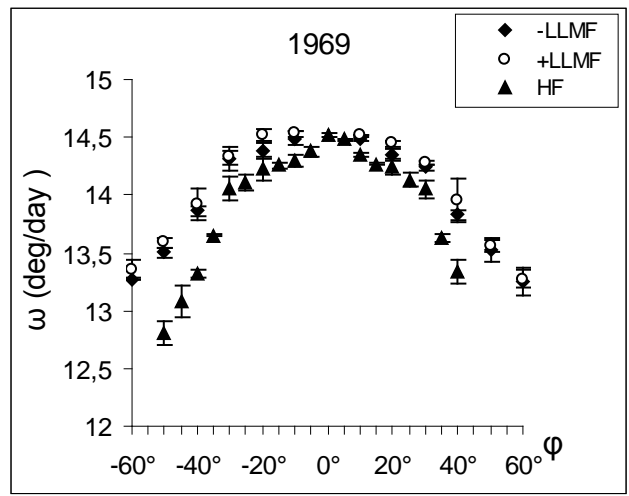
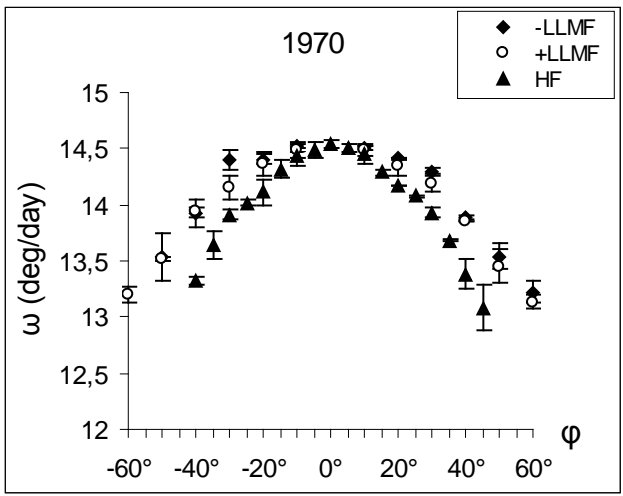
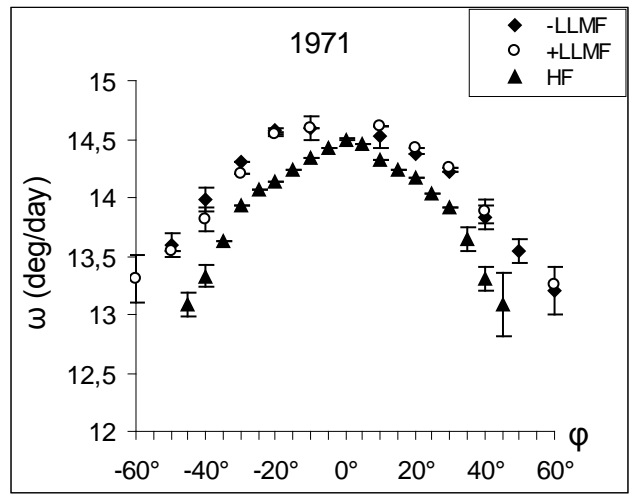
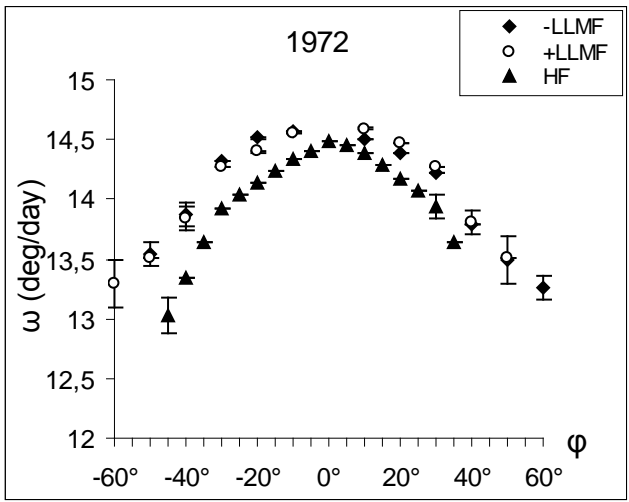
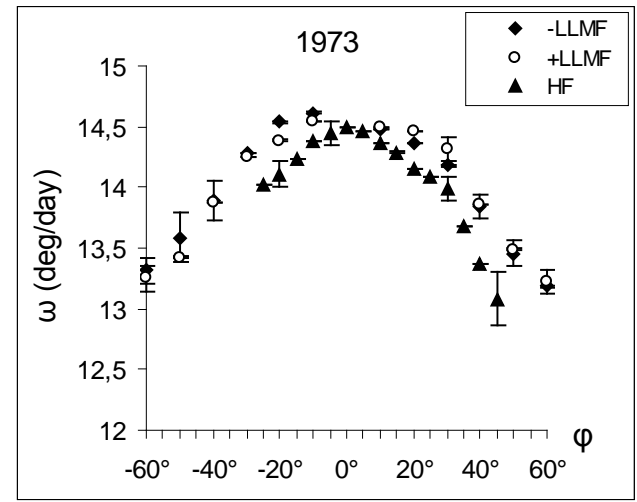

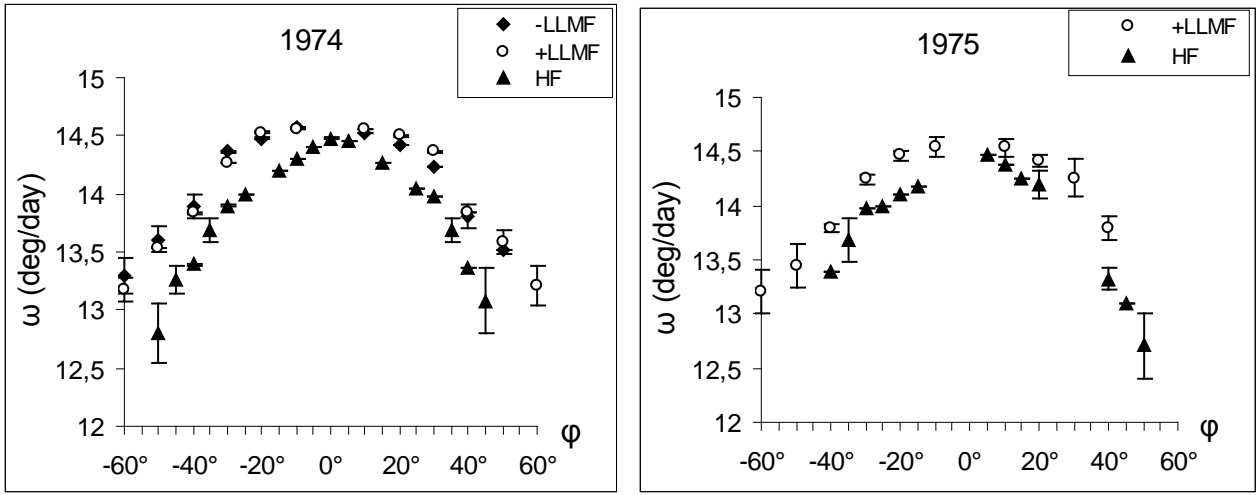

Figure 3. Rotation rates of Hα filaments (HF) and of long-lived magnetic features of positive (+LLMF) and negative polarity (-LLMF) for the Solar Cycle 20. Solar latitudes (φ) are presented on the abscissas and rotation rates (ω) - on the ordinates.

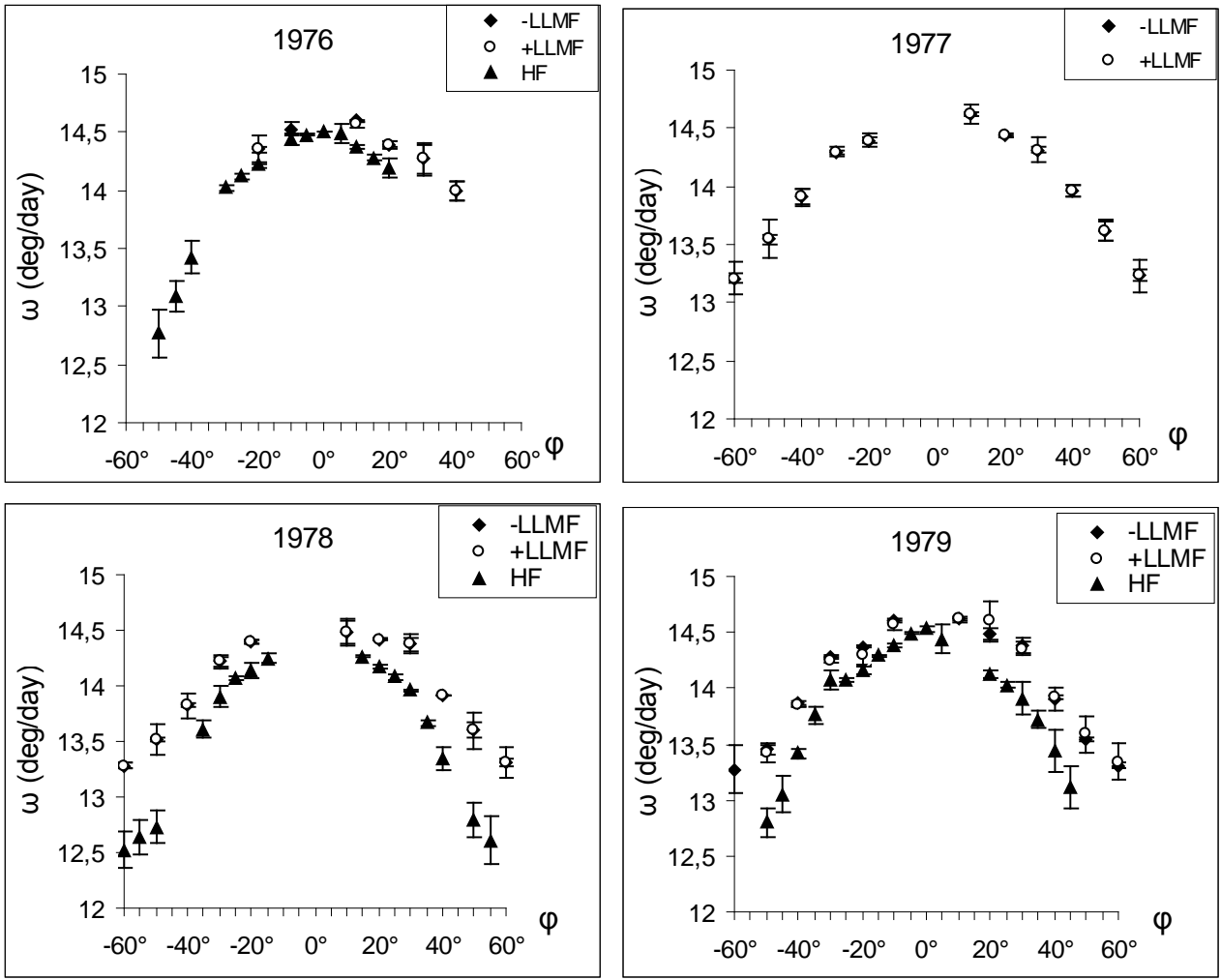

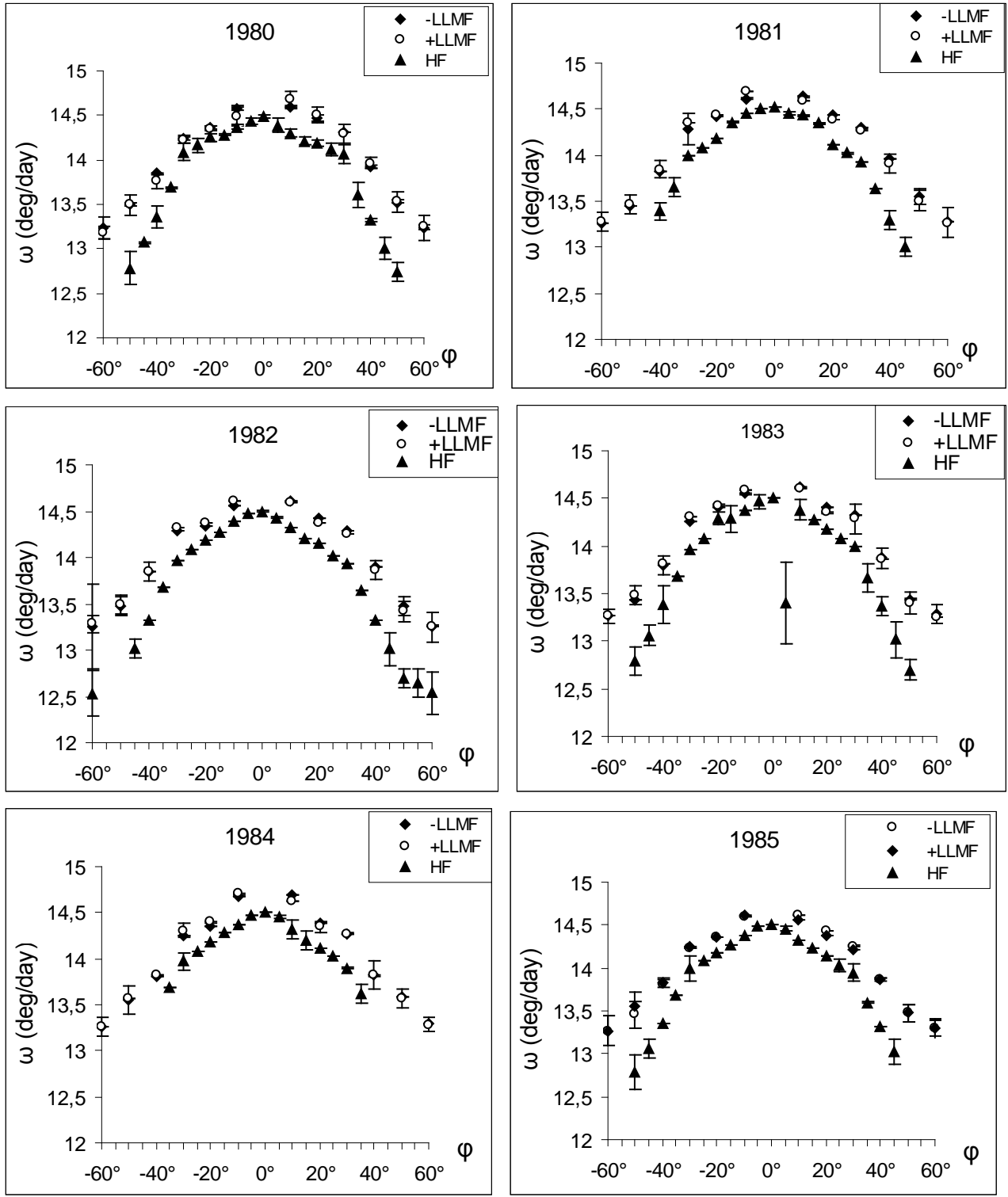

Figure 4. The same as in the Figure 3 for Solar Cycle 21.

From the Figures (3, 4) we can see that the long-lived magnetic features have higher rotation rates than Hα filaments during Solar Cycles 20 and 21. The figures clearly indicate that long-lived magnetic features of negative and positive polarity show more rigid rotation than Hα filaments.

Using least-squares fitting the Faye formula was derived for differential rotation of Hα filaments and long-lived magnetic features. For Hα filaments we have obtained ω=14.49(±0.04)–2.70(±0.09)

$\sin^2\varphi$ and $\omega=14.49(\pm0.04)-2.73(\pm0.09) \sin^2\varphi$ for Solar Cycles 20 and 21 respectively. Similarly, for the long-lived magnetic features we obtained $\omega=14.66(\pm0.04)-1.90(\pm0.09) \sin^2\varphi$ and $\omega=14.67-1.89(\pm0.09) \sin^2\varphi$ for Solar Cycles 20 and 21, respectively.

By the comparative analyses of observational data of Hα filaments and long-lived magnetic features for the solar activity Cycles 20 and 21 we have found that the Hα filaments have lower rotation rates and they rotate more differentially. The long-lived magnetic features have higher rotation rates and rotate more rigidly than Hα filaments.

## 5. Discussion and Conclusions

Quiescent filaments are aligned along inversion lines of the large-scale magnetic field, and thus they reveal the rotation of global magnetic field patterns. Therefore, it is important to use extensive data sets covering long observation periods to investigate them. In particular, continuous observations of Hα quiescent filaments can be useful for studying the variations of global circulation over a broad range of heliographic latitudes in the solar atmosphere.

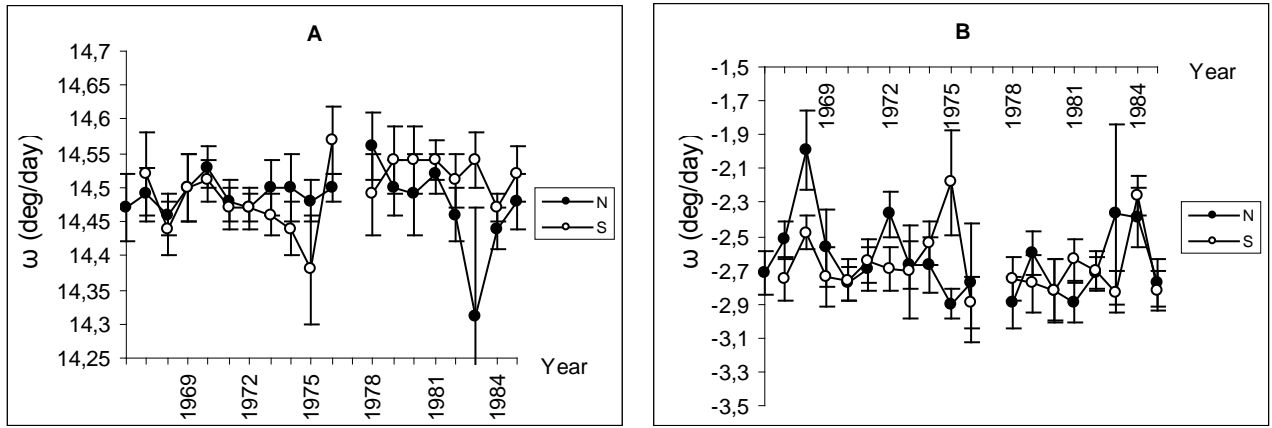

Figure 5. Temporal variation of equatorial angular velocity (A) and the latitude gradient coefficient (B) of rotation rate (ω) of Hα filaments for the two solar hemispheres during Solar Cycles 20 and 21.

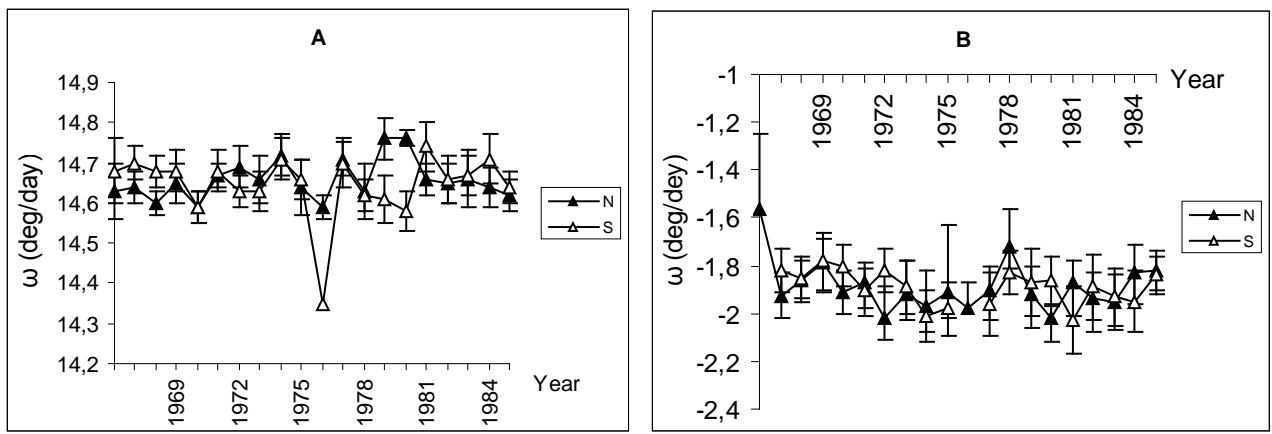

Figure 6. Temporal variation of equatorial angular velocity (A) and the latitude gradient coefficient (B) of rotation rate (ω) of long-lived magnetic features of positive polarity for two solar hemispheres during Solar Cycles 20 and 21.

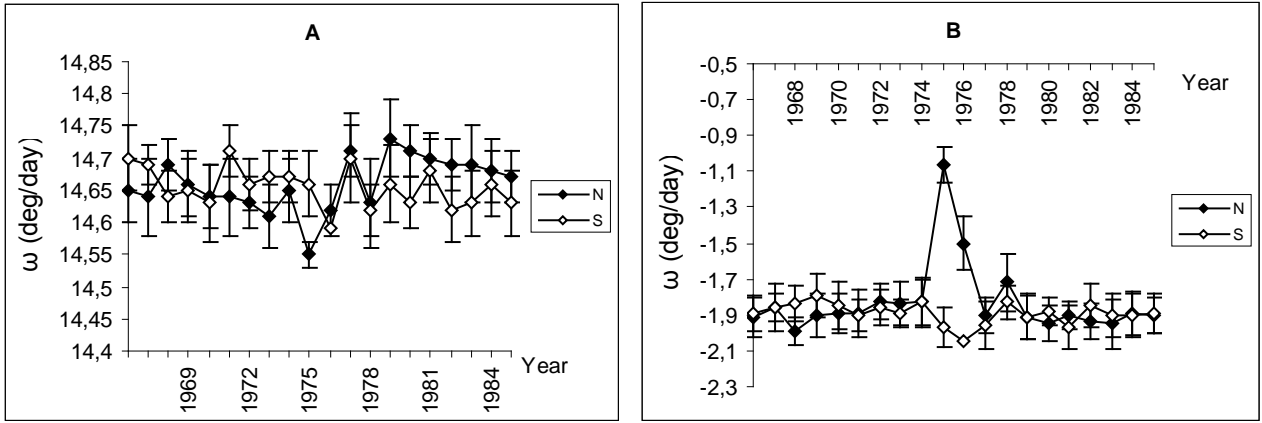

Figure 7. Temporal variation of equatorial angular velocity (A) and the latitude gradient coefficient (B) of rotation rate (ω) of long-lived magnetic features with the negative polarity for two solar hemispheres during Solar Cycles 20 and 21.

It is clear from the Figures 5-7, that faster rotation rates of Hα filaments is observed around sunspot minima for different solar hemispheres for equatorial angular velocity (A), as well as for the latitude gradient coefficient (B). Near the minimum of solar cycle (in 1976) equatorial rotation rates of the long-lived magnetic features of positive polarity are minimal for both solar hemispheres during Solar Cycle 20 as well as in 1985 (a year before minimum), where exists a tendency of decreasing rotation rate. Similar variations are observed for long-lived magnetic features of negative polarity in 1975-1976 and 1985.

The small homogeneous variations are characteristic for the coefficient B of the long-lived magnetic features with the positive polarity for both hemispheres for Solar Cycles 20 and 21. However, for the long-lived magnetic features with the negative polarity, coefficient B revealed N-S asymmetry and significant amplitude variation in Northern hemisphere near the minimum (in 1975-1976).

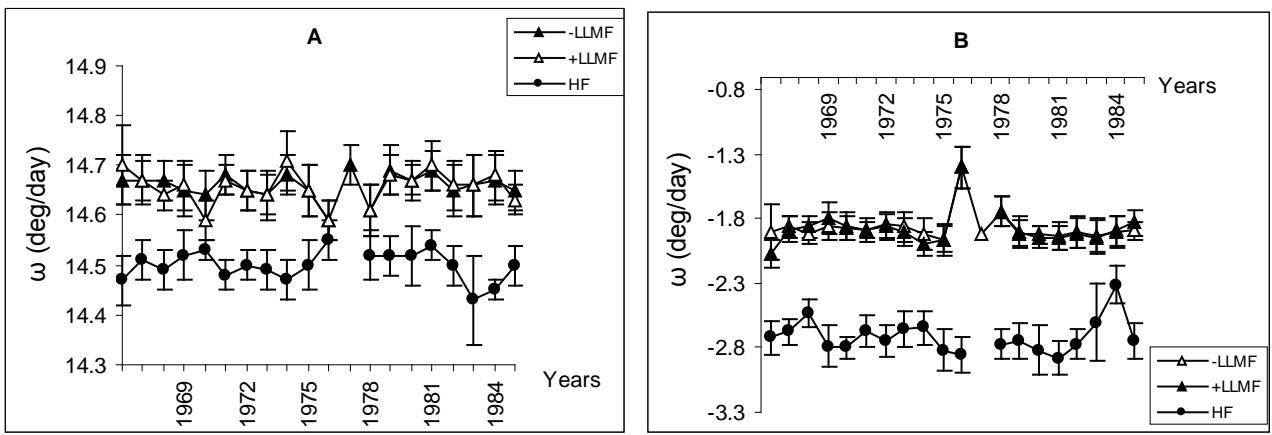

Figure 8. Temporal variation of equatorial angular velocity (A) and the latitude gradient coefficient (B) of rotation rate (ω) for the solar features analyzes averaged over the hemispheres during Solar Cycles 20 and 21

Variations of the equatorial rotation rates (A) and the latitude gradient coefficient (B) for Hα filaments and the long-lived magnetic features with the negative and positive polarities are presented for the Solar Cycles 20 and 21 in Figure 8. It is possible to see that Hα filaments and the long-lived magnetic features reveal asymmetry in rotation rate for investigated period (1966-1985). The N-S asymmetry of rotation rates needs further studies and it will be the subject of our future investigations, carried out by improved statistical investigation, than we used for Hα filaments in earlier works (Gigolashvili et al., 2005; 2007).

In the present work we find that the rotation rate on the equator (A), as well as the latitude gradient coefficient (B) shows a systematic variation within each cycle. Near the cycle minima the Hα filaments and the long-lived magnetic features of negative and positive polarities have higher rotation rates. This result agrees with the result of Balthasar, Vazquez, and Woehl (1986). They found the highest velocity around the minima and that at the start of an activity maximum there is a secondary velocity maximum.

In the followings we compare our results with results of other authors'. In 1919-1930, based on observations of 105 long-lived filaments in the $Ca^+$ $K_3$ line, D'Azambuja and D'Azambuja (1948) obtained the Faye formula: $\omega(\varphi) = 14.48 - 2.16 \sin^2\varphi$ deg day$^{-1}$. For the long-lived filaments, Glackin (1974) obtained the Faye formula: $\omega(\varphi) = 14.51 - 2.16 \sin^2\varphi$ deg day$^{-1}$. The latter results were derived by the observation of eight stable filaments following them for one or more solar rotations during 1971-1972.

Deng, Wang, and Harvey (1999) have shown that polar magnetic elements could live from several to more than 58 hours. This enables to realize measurements of the solar rotation rate near the polar region by tracing magnetic element motions. By two fitting methods they have obtained $\omega = 14.0 \pm 0.54 - (2.24 \pm 1.22) \sin^2\varphi - (1.78 \pm 0.79) \sin^4\varphi$ deg day$^{-1}$ for the mean sidereal rotation rate.

Using the least-squares method we have found coefficients of Faye's formulas for Hα filaments and long-lived magnetic features with negative and positive polarities (Equations (1) and (2), accordingly). Hα filaments rotation rates are in the range of earlier results by D'Azambuja and D'Azambuja (1948) and Glackin (1974).

Based on the data set of sunspots for three solar cycles Kambry and Nishikawa (1990) have found that the differential rotation varies from cycle to cycle. They revealed that the rotation velocity in a low-activity cycle (Cycle 20) is higher than in a high-activity cycle (Cycle 19). On the contrary, Javaraiah (2003) using Greenwich data (1879-1976) and SOON/NOAA data (1977-2002) of sunspot groups found that the Sun's mean equatorial rotation rate (A) is significantly larger (≈0.1 %) in the odd-numbered sunspot cycles than in the even-numbered sunspot cycles.

Using the solar full disc observation of Nobeyama Radioheliograph at 17 GHz for the period 1999-2001 Chandra, Vats, and Iyer (2009) have shown that the solar corona rotates less differentially than the photosphere and chromosphere.

Kariyappa (2008) had also used the X-ray Bright Points (XBPs) as tracers for determining the coronal rotation. The analysis of the XBPs over the solar disk have clearly indicated that the solar corona rotates differentially like the photosphere and chromosphere, and the sidereal angular rotation velocity is independent of the sizes of the XBP used as a tracer.

Hara (2009) statistically estimated the differential rotation rate of the solar corona from the motion of XBPs. There were observed with the Yohkoh soft X-ray telescope and the source region of magnetic fields is discussed from the evaluated rotation rate. The differential rotation rate shows a similar trend to that of photospheric magnetic fields. The differential rotation rate changes with a parameter that is associated with the lifetime of XBPs, and that it becomes smaller with height. In the Author's opinion this trend suggests that magnetic fields associated with XBPs with a short lifetime are rooted just below the surface of the Sun at the top of the convection zone, and that they have a different origin from active regions (Hara, 2009). However, Weber et al. (1999) and Weber and Sturrock (2002) studied the rotation rate of corona by using Yohkoh/SXT data and showed that the soft X-ray corona do not exhibit any significant differential rotation. It means that the corona rotates more rigidly than the chromosphere and the photosphere.

Investigations of the latitudinal variation in the solar rotation in soft X-ray corona using Yohkoh/SXT data (Chandra, Vats, and Iyer, 2010) show that the equatorial rotation rates vary systematically with sunspot numbers, indicating a dependence on the phases of the solar activity cycle.

D'Azambuja and D'Azambuja (1948) and Glackin (1974), as well the present work have derived higher rotation rates for filaments than sunspots rotation rates obtained by other authors (Newton and Nunn, 1951; Kambry and Nashikawa, 1990).

Coefficients A and B for Eqation (1) for different solar features obtained by various authors are presented in Table 2.

Table 2. The coefficients A & B of solar different rotation obtained by various authors

| Years, cycle number and cycle phase of observation | Coefficients $A \pm \Delta A$ | $B \pm \Delta B$ | Tracers | Source |
|---|---|---|---|---|
| 1934-1944 (Cycle 17) | 14.38±0.01 | -2.96 ± 0.09 | Recurrent sunspots | Newton and Nunn (1951) |
| 1955-1965 (Cycle 19) | 14.38 ± 0.02 | -2.65 ± 0.16 | Sunspots | Kambry and Nashikawa (1990) |
| 1966-1975 (Cycle 20) | 14.52 ± 0.02 | -2.53 ± 0.22 | | |
| 1976-1985 (Cycle 21) | 14.44 ± 0.02 | -2.22 ± 0.18 | | |
| 1967-1987(Cycles 20, 21) | 14.71±0.01 | 2.39 ±0.04 | Solar plasma | Snodgrass (1990) |
| 1919-1930 (descending phase of Cycle 15, minimum, ascending and maximum phases of Cycle 16) | 14.48 | -2.16 | Long-term filaments | D'Azambuja and D'Azambuja (1948) |
| 1971-1972 (descending phase of Cycle 20) | 14.51 | -2.16 | Filaments | Glackin (1974) |
| 1966-1975 (Cycle 20) | 14.49 ± 0.04 | -2.70 ± 0.09 | Filaments | Present Work |
| 1976-1985 (Cycle 21) | 14.49 ± 0.04 | -2.73 ± 0.09 | | |
| 1966-1975 (Cycle 20) | 14.66 ± 0.04 | -1.90 ± 0.09 | Long-lived magnetic features | Present Work |
| 1976-1985 (Cycle 21) | 14.67 ± 0.04 | -1.89 ± 0.09 | | |
| 8-15 July 1997 (the minimum between Cycles 22 and 23) | 14.0±0.54 | -2.24±1.22 | Near the polar region magnetic element | Deng et al., 1999 |
| March and April, 2007 (the minimum between Cycles 23 and 24) | 14.19 ± 0.04 | - 4.21 ± 0.78 | XBPs | Kariyappa, 2008 |
| 4 January 1994, 27 December 1997 (the minimum between Cycles 22 and 23) | 14.39±0.01 | -1.91±0.10 | XBPs | Hara, 2009 |
| 1999-2001 (the maximum of Cycle 23) | 14.82±0.06 | -2.13±0.14 | Coronal Radio Emission (17 GHz) | Chandra et al., 2009 |
| 1992-2001 (descending phase of Cycle 22, minimum and ascending phases of Cycle 23) | 14.5±0.1 | -0.8±0.6 | X-Ray Corona | Chandra et al., 2010 |

Table 2 indicates that the A and B coefficients are different in different solar activity cycles and for different solar features.

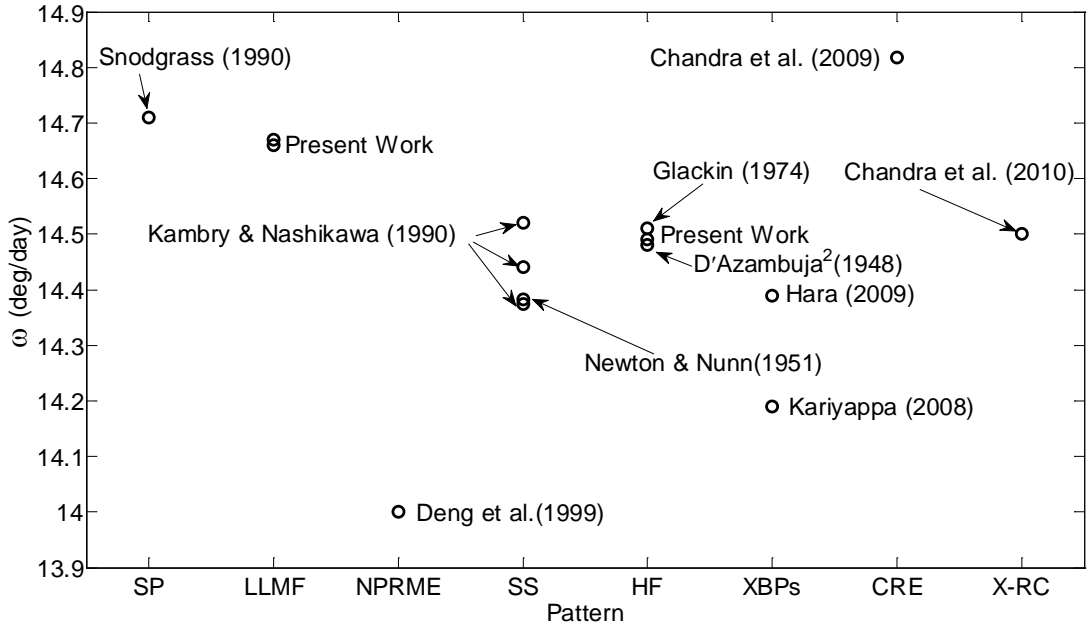

Figure 9. Summarizing data of various solar features' equatorial rotations rates (ω) ordering them according to their approximate height in the solar atmosphere. These features are: SP – Solar Plasma, LLMF – long-lived magnetic features, NPRME - Near the Polar Region Magnetic Element, SS – Sunspot, HF – Hα filament, XBPs - X-ray Bright Points, CRE - Coronal Radio Emission (17 GHz), and X-RC - X-Ray Corona.

As we can see from the Figure 9 and Table 2, the equatorial rotation rates (A) for various features of the solar surface vary in a wide range from 14.0±0.54 (Deng et al., 1999) to 14.817±0.060 (Chandra et al., 2009). The coefficient B also varies from 0.8±0.6 (Chandra et al., 2010) for the X-Ray Corona to 2.96 ± 0.09 (Newton and Nunn (1951) for recurrent sunspots.

Rotation rates of filaments received by us are larger by ~0.07% than rotation rates derived by D'Azambuja and D'Azambuja (1948) and by ~0.14% lower than rotation rates obtained by Glackin (1974). There is a very good agreement between these results by different authors. This is possible in the case when precise measurements of well-identified filaments in consecutive days are carried out. The rotation rates of Hα filaments obtained by us are near to the rotation rates of X-Ray Corona's that were derived by Chandra, Vats, and Iyer (2010), too. Namely, the X-Ray Corona's rotation rate is faster than that of filaments by only about 0.07%. As to velocities of long-lived magnetic features, measured by us, they are near to values of solar plasma' velocities (Snodgrass and Ulrich (1990), i.e. lower by ~0.31%. Nevertheless, differences between rotation rates of the long-lived magnetic features and magnetic features near to poles are much larger – about 4.54%. The latter can be explained by the difficulties of measurements at high latitudes.

Rotation rate derived from radio emission at 17 GHz and that of the X-Ray Corona by Chandra et al. (2009, 2010) are by ~1.04% larger and by ~1.13% lower respectively, than velocities of the long-lived magnetic features, measured by us in present work. LLMFs' rates are by ~1.5-2% higher than rotation rates of sunspots (Newton and Nunn, 1951; Kambry and Nashikawa, 1990), and by ~2% higher than rotation rates of XBPs obtained by Hara (2009).

Very high discrepancies in the values of rotation rates, derived for the X-Ray Corona (Chandra et al., 2010) and XBP (Kariyappa, 2008; Hara, 2009) may arise from difficulties in definition of parameters that are associated with the lifetime and height of XBPs in the solar atmosphere.

By analyzing of the differential rotation rates of various features in the solar atmosphere for the Solar Cycles 20 and 21, we have established that in this period the Hα filaments had lower rotation rates and rotated more differentially than long-lived magnetic features. Near the solar activity cycle minima the Hα filaments and the long-lived magnetic features of negative and positive polarities had higher rotation rates than during the maxima of the cycles. The relationship between the rotation rate of Hα filaments and the photospheric small-scale magnetic structures is important for understanding the nature of solar activity. Therefore, it is important to consider in any theory the origin and evolution of solar magnetic fields their connection with processes in the solar interior.

The rigid component in the rotation rate of magnetic features, found in this paper, might be consistent with the long lifetime of some large-scale patterns formed by low temperature regions observed in the millimeter wavelength range (Vrsnak et al., 1992). These radio structures are highly correlated with magnetic inversion lines. The rigid rotation of the magnetic features on the Sun were also reported by Bumba and Hejna (1987), Sheeley, Nash, and Wang (1987) and by Stenflo (1989). However, there is no unique interpretation of the rigid rotation component. So, Sheeley, Nash, and Wang (1987) proposed a magnetic flux-transport model, while Stenflo (1989) proposed a replenishment of surface magnetic flux by sources from the convection zone.

Finally, we have found indications of a higher rotation velocity during the minimum phases of the solar magnetic cycle. This is consistent with some theoretical models (e.g., Brun, 2004; Lanza, 2007).

**Acknowledgements**

We are grateful to authors of the "Atlas of Stackplots" (1991), published by World Data Center A for Solar-Terrestrial Physics National Geophysical Data Center. We thank also unknown referees for useful critical remarks.